\documentclass[prd,amsmath,notitlepage,twocolumn,nofootinbib,superscriptaddress]{revtex4-1}
\usepackage{graphicx}
\usepackage{float}
\usepackage{epstopdf,cancel}
\usepackage{epsf,latexsym,bbm,euscript}
\usepackage{amssymb,amsmath}
\usepackage{mathtools} 
\usepackage{times,graphics}
\usepackage{soul,xcolor}

\def\beq{\begin{equation}}
\def\eeq{\end{equation}}

\def\be{\begin{equation}}
\def\ee{\end{equation}}

\def\6{\langle}
\def\9{\rangle}

\def\bali{\begin{align}}
\def\eni{{\end{align}}}

\def\va{{\vec{a}}}

\newcommand\vom{{\vec{\omega}}}

\newcommand\vsi{{\vec{\sigma}}}

\def\hn{{\hat{n}}}
\def\half{\tfrac{1}{2}}

\def\vbe{{\vec{\beta}}}

\newcommand{\defeq}{\vcentcolon=}

\begin{document}

\title{Quantum satellites and tests of relativity}\let\thefootnote\relax\footnotetext{Submitted to the Proceedings of the Fifteenth Marcel Grossmann Meeting, Rome, July 2018\\} 

\author{Piergiovanni~Magnani} \address{{Department of Physics, Politecnico di Milano, Milano 20133, Italy}}
 \author{Matteo~Schiavon}
\address{Dipartimento di Ingegneria dell'Informazione, Universit\`{a} degli Studi di Padova, Padova 35131, Italy}
 \address{Istituto Nazionale di Fisica Nucleare (INFN) --- Sezione di Padova, Italy}  
 
 \author{Alexander~R.~H.~Smith} \address{Department of Physics and Astronomy, Dartmouth College, Hanover, New Hampshire 03755, USA}
  \author{Daniel R. Terno}
  \email{daniel.terno@mq.edu.au}  
\address{Department of Physics and Astronomy, Macquarie University, Sydney, NSW 2109, Australia}  

\author{Giuseppe~Vallone}
\author{Francesco~Vedovato}
\author{Paolo~Villoresi } \address{Dipartimento di Ingegneria dell'Informazione, Universit\`{a} degli Studi di Padova, Padova 35131, Italy}
 \address{Istituto Nazionale di Fisica Nucleare (INFN) --- Sezione di Padova, Italy} \author{Sai Vinjanampathy}
 \address{Department of Physics, Indian Institute of Technology-Bombay, Powai, Mumbai 400076, India}   \address{Centre for Quantum Technologies, National University of Singapore, 3 Science Drive 2, 117543 Singapore, Singapore}
  
 \begin{abstract} Deployment of quantum technology in space provides opportunities for new types of precision tests of gravity. On the other hand, the operational demands {of such technology}
 can make previously unimportant effects practically relevant.
 We describe a {novel} optical interferometric red-shift measurement and {a measurement scheme designed to witness   possible spin-gravity coupling effects.}

\end{abstract} \maketitle
\section{Introduction}

Amazing experimental progress  in quantum sensing
and quantum communications together with satellite deployment of quantum
technologies have ushered in a new era of experimental physics in outer space.

The success of the first space based quantum key distribution experiments performed with the Micius satellite \cite{chinasat} is expected to be soon followed by European and North American missions.   At the same time current  missions, such as LAGEOS-2, BEACON-C and LCT on Alphasat I-XL, are adapted for
quantum optics experiments \cite{padova1,german}.
While the primary goal of these space-based platforms is to provide links for global quantum key distribution, the missions also envisage substantial scientific programs.
These experiments have the exciting potential to open up new tests of fundamental physics by enabling new searches for signatures of quantum gravity and/or physics beyond the standard model \cite{qsat}.
On the other hand, the ambitious precision and stability goals \cite{qclock} are likely to turn the questions of gravitational and inertial effects on spin into practical questions.

Here we describe how these technologies can be affected and used to test the  Einstein Equivalence Principle (EEP). The principle comprises three statements \cite{will:lrr,will:93,wp:book}.
 The first --- {\it Weak Equivalence Principle}  (WEP) --- states that the trajectory of a freely falling test body is independent of its internal composition.
  Closely related to {the WEP} is the Einstein elevator:
 if all bodies fall with the same acceleration
in an external gravitational field, then to an observer in a small freely falling
lab in the same gravitational field, they appear unaccelerated~\cite{will:93}.
{The remaining} two statements deal with outcomes of non-gravitational experiments performed in freely falling laboratories where self-gravitational effects are negligible.
 The second statement --- {\it Local Lorentz Invariance} --- asserts that such experiments are independent of the velocity of the laboratory where the experiment takes place.
  The third statement --- {\it Local Position Invariance} (LPI) --- asserts that ``the outcome of any local non-gravitational experiment is independent of where and when in the universe it is performed''~\cite{will:lrr}.

In Sec.~II we outline a novel all-optical test of  LPI. In Sec.~III we discuss the inertial and  aspects of the spin{-gravity} coupling and suggest the  weak valued amplification scheme for detecting some of these effects.
          \section{Optical test of position invariance}

Tests of the ``when'' part of  LPI bound the variability of the non-gravitational constants over  cosmological time scales~\cite{uzan:lrr}.
The ``where'' part  was expressed in Einstein's analysis of what in modern terms is a comparison of two identical frequency
standards in two different locations in a static gravitational field. The so-called {\it red-shift} implied by the LPI affects the locally measured
 frequencies of a spectral line that is emitted at location $1$ with $\omega_{11}$ and then detected at location $2$ with $\omega_{12}$.
The red-shift can be parameterized at the leading post-Newtonian order as
\be
\Delta \omega / \omega_{11} = (1 + \alpha) (U_2 - U_1) + \mathcal{O}(c^{-3}),
\ee where $\Delta \omega := \omega_{12} - \omega_{11}$, $U_i:=-\phi_i/c^2$ has the opposite sign of the Newtonian gravitational potential $\phi_i$ at the emission ($1$) and detection ($2$),
 while $\alpha \neq 0$ accounts for possible violations of LPI.
In principle, $\alpha$ may depend on the nature of the clock that is used to measure the red-shift~\cite{will:lrr}. The standard model extension   includes variously constrained parameters that
predict LPI violation \cite{sme,lli}. Alternative theories of gravity not ruled out by
current data also predict $\alpha\neq 0$~\cite{will:lrr,altsat}.

 A typical red-shift experiment  involves a pair of clocks, naturally occurring~\cite{sunline} or specially-designed~\cite{pr:60,gpa:80,app:18},
  whose readings are communicated by electromagnetic  (EM) radiation.
   The resulting estimates of $\alpha$ are based on comparison of fermion-based standards. Hence, different types of experiments, which employ a single EM-source and compare optical phase differences
      between beams of light traversing different paths in a gravitational field, provide a complementary test of LPI.

Such an all-optical  experiment  was proposed   as a possible component
 of the QEYSSAT mission~\cite{qsat}. A photon time-bin superposition is sent from a ground station on Earth to a spacecraft,
  both equipped with an interferometer of imbalance $l$, in order to temporally recombine the two time-bins and obtain an interference pattern depending on the gravitational phase-shift:
\begin{equation}
\varphi_{\rm gr} = \frac{\Delta \omega}{\omega} \frac{2\pi}{\lambda} l \approx (1+\alpha) \frac{2\pi}{\lambda}\frac{gh l }{c^2} \ ,
\label{phigr}
\end{equation}
where $g$ is the Earth's gravity, $h$ the satellite altitude and $\lambda = 2\pi c/\omega$ the sent wavelength.
For $\alpha = 0$, this phase-shift is of the order of few radians supposing $l = 6$~km, $\lambda = 800$~nm and $h = 400$~km~\cite{qsat}.

However, a careful analysis of this optical COW-like experiment~\cite{bggst:15} revealed that first-order Doppler effects are roughly $10^5$ times
stronger than the desired signal $\varphi_{\rm gr}$ from which $\alpha$ would be estimated. This first-order Doppler effect was recently
measured by exploiting large-distance precision interferometry along space channels~\cite{padova:16}, which constitute a resource for performing
fundamental tests of quantum mechanics in space and space-based quantum cryptography.

We propose \cite{sat-us} a new gravitational red-shift experiment, which uses a single EM-source and a double
large-distance interferometric measurement performed at two different gravitational potentials. By comparing the phase-shifts obtained at a
satellite and on Earth, it is possible to cancel the first-order Doppler effect. Thus, this experimental proposal allows for a bound on $\alpha$
quantifying the violation of LPI in the EM-sector with a precision on the order of $10^{-5}$.

\begin{figure}[h]
\begin{center}
\includegraphics[width=3.6in]{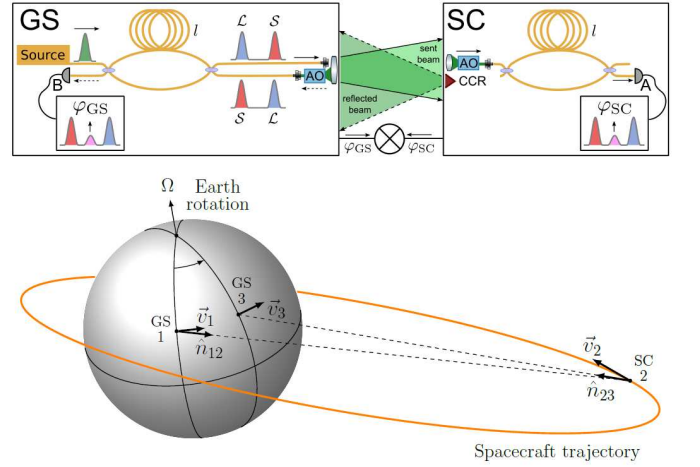}
\end{center}
\caption{Top: A schematic diagram of the proposed experiment.
 Both the ground station (GS) and spacecraft (SC) are equipped with a MZI of equal delay line $l$ and an adaptive optics system for fibre injection.
  Bottom: The geometry of the GS and SC used in the experiment, where $\vec{v}_1$ is the velocity of the GS at the emission location and potential $U_1$;
  $\vec{v}_2$ is the velocity of the SC at the detection location on the satellite and potential $U_2$; $\vec{v}_3$ is the velocity of the ground station at the detection
  of the beam retro-reflected by the SC, which occurs at a potential $U_3=U_1$. }
\label{aba:fig1}
\end{figure}

This proposal \cite{padova:16} is comprised of an interferometric measurement obtained by sending a
light pulse through a cascade of two fiber-based Mach-Zehnder interferometers (MZI) of equal temporal
imbalance $\tau_l$. After the first MZI, the pulse is split into two temporal modes, called {\it short} ($\mathcal{S}$) and {\it long}
 ($\mathcal{L}$) depending on the path taken in the first MZI. The equal imbalance of the two MZIs guarantees that the two pulses are
 recombined at the output of the second MZI, where they are detected.  Such a satellite interferometry experiment setup is sketched in \cite{aba:fig1}.

The combination of the possible paths the pulses may take leads to a characteristic detection pattern comprised of three
possible arrival times for each pulse, as depicted in the insets of the upper picture in Fig.~\ref{aba:fig1}. The first
(third) peak corresponds to the pulses that took the $\mathcal{S}$ ($\mathcal{L}$) path in both the MZIs, while the mid
 peak is due to the pulse that took the $\mathcal{S}$ path in the first interferometer and the $\mathcal{L}$ path in the
  second interferometer, or vice versa. Hence, interference is expected only in the central peak due to the indistinguishability of these
   latter two possibilities. A successful realization of the experiment depends on a number of important technical aspects that are described in detail in \cite{sat-us}.

A bound on $\alpha$ is retrieved from the difference of the two phase-shifts, $\varphi_{\rm SC}$ and $\varphi_{\rm GS}$, that are
obtained from interferometric measurements on the spacecraft and ground station, respectively. As just described, the interfering beams take different paths in the passage through the two MZIs.
At the satellite, the beam that took the $\mathcal{L}$ path on Earth and the $\mathcal{S}$ path on the spacecraft interferes with the beam that passed took the $\mathcal{S}$ path
on Earth and then took the $\mathcal{L}$ one on the spacecraft. This interference is a result of the phase difference $\varphi_{\rm SC}$. Analogously, at the ground station (GS)
 the beams that were delayed on the Earth before and after their round trip to the spacecraft (SC) will also interfere because of the phase difference $\varphi_{\rm GS}$.

The signal from which a bound on $\alpha$ is obtained is a linear combination of the two measured phase-shifts
\be
 \varphi_{\rm SC}= (\omega_{12}-\omega_{11})\tau_l,  \quad    \varphi_{\rm GS}=  (\omega_{13}-\omega_{11})\tau_l,
 \ee
where $\omega_{11}$ is the proper central frequency  of the emitted signal at the GS and $\omega_{13}$ is the frequency after the round trip, and the proper delay time
$\tau_l$ is the same in both frames.

The standard second-order expression for the frequencies detected at the satellite   is
\be
 \frac{\omega_{12}}{\omega_0} = {\left( \frac{1-U_1-\half \beta_1^2}{1-U_2-\half \beta_2^{2}} \right) \left( \frac{1-\hn_{12}\cdot\vbe_2 }{1-\hn_{12}\cdot\vbe_1 } \right)},
 \ee
and at the ground station after a go-return trip 
\be
  \frac{\omega_{13}}{\omega_0}= \left(\frac{1-\hn_{23} \cdot \vbe_3 }{1-\hn_{23}\cdot\vbe_2 }\right)\left(\frac{1-\hn_{12}\cdot \vbe_2 }{1-\hn_{12}\cdot\vbe_1}\right),
 \ee
 where $\vbe_i\defeq\vec{v}/c$.
The first-order Doppler terms are eliminated by manipulating the corresponding data sets from the GS and SC in a manner similar to  time-delay interferometry
techniques~\cite{tdi} and those used in the Gravity Probe A experiment \cite{gpa:80}. The key feature allowing for
this elimination is that the ratio of the  first-order Doppler effect contributions to the two signals, $\varphi_{\rm SC}$ and $\varphi_{\rm GS}$, is exactly two \cite{sat-us}. Hence the target signal is
\be
S\defeq \varphi_{\rm SC}-\tfrac{1}{2}\varphi_{\rm GS},
\ee
leading to
\begin{align} \frac{S}{\omega_0 \tau_l}& =(1+\alpha) (U_2-U_1)\nonumber \\
&+\tfrac{1}{2}(\vec{\beta}_1-\vec{\beta}_2)^2  -(\mathfrak{d}_1 -\mathfrak{d}_2)^2 -  {T} \hat{n}_{12}\cdot\va_1/c, \label{eq_Sloop}
\end{align}
where $\vec{\beta_i}=\vec{v}_i/c$, $\mathfrak{d}_i=\hat{n}_{12}\cdot\vec{\beta_i}$, $\va_1$ is the centripetal acceleration at the GS, and $T$ is the upward propagation time.

       \section{Weak equivalence principle and orbiting clocks}

Matter of the Standard Model is characterized by two parameters of the irreducible representations of
the Poincar\'{e} group: mass and spin (or helicity). General relativity is a universal interaction theory about masses \cite{ni:10}, like the Newtonian gravity, with polarization
effects implicitly omitted from the WEP. Precision measurements up-to-date have not revealed spin-gravity coupling, but it is clearly conceivable \cite{ni:10}.

  Regardless of their origins,   spin-gravity coupling terms provide effective corrections to the  Hamiltonian in the limit of weak gravity  and non-relativistic motion.
   The leading terms of the Hamiltonian of a free spin-$\frac{1}{2}$ particle
that take into account the effects of  rotation of the reference frame with  angular velocity $\vec\omega$ and   acceleration $\vec a$ (or a uniform gravitational field)
can be represented as
\be
H=H_\mathrm{cl}+H_\mathrm{rel}+H_\sigma+H_\mathrm{ext}.
\ee
 The first three terms on the right hand side are obtained by performing the    standard Foldy-Wouthuysen transformation and taking the non-relativistic limit~\cite{hn:90}. The term
$ H_\mathrm{cl}$ represents the standard Hamiltonian of a free non-relativistic particle in a non-inertial frame,  $H_\mathrm{rel}$ describes the higher-order relativistic corrections
 that do not involve spin,    and
  \be
  H_\sigma=-\tfrac{1}{2}\hbar\vom\cdot \vsi  +\frac{\hbar}{4mc^2}\vec\sigma\cdot(\va\times \vec p).
   \ee
 Finally, the term
\be\label{PO}
H_\mathrm{ext}=\frac{\hbar k}{2c}\va\cdot\vsi,
\ee
represents the spin-accelerating (or spin-gravity) coupling.  It is a limiting form of the simplest phenomenological addition to the Dirac equation that breaks the WEP \cite{p78}.
For the value $k=1$       it results from a particular version of the  Foldy-Wouthuysen  transformation \cite{obu}. While commonly considered a mathematical artefact of this transformation,
the term naturally arises in  gravitational{ly inspired} Standard Model extensions.

The Mashhoon term $-\tfrac{1}{2}\hbar\vom\cdot \vsi$ was recently detected by using neutron polarimetry \cite{neut}. On the other hand,
 only model-dependent bounds on $k$ in  $H_\mathrm{ext}$ were obtained by a variety of techniques \cite{rmp:18}, including the optical magnetometery \cite{kimball:13}.

The spin dependent terms are small under normal conditions.
On the Earth's surface $\hbar g/c=2.15\times 10^{-23}$ eV, which is equivalent to {an} effective magnetic field of $3.7\times 10^{-19}$ Tl, still several orders of magnitude
below the peak sensitivity of  optical magnetometery.     The spin-rotation term is significantly larger, since already on the ground $\omega c/g=2.22\times 10^3$.
It will be about an order of magnitude stronger for low-orbit satellites that are planned to carry entangled optical clocks \cite{qclock} aiming to establish the  precision of $10^{-18} - 10^{-20}$,
making it a factor to consider {in} the clock design.

A potentially promising way of detecting these effects is via so-called weak amplification~\cite{weak-m}.
 Weak value amplification involves two systems (typically referred to as ``system'' and ``meter'') that can interact via an interaction Hamiltonian of the form $q\delta(t-t_0)\hat{A}\otimes\hat{p}$.
 The bipartite system-meter is prepared in an initial state $|{s_i}\rangle\otimes|{m_i}\rangle$, following which the two are allowed to interact for a small time that includes $t_0$.
  Following this, the system is measured and measurements corresponding to a post-selected system state $|{s_f}\9$ are considered. This pre- and post-selection induces a ``kick'' in the meter state,
   given by the evolution $e^{-iq \mathcal{A}_w \hat{p}}|{m_i}\rangle$, where $\mathcal{A}_w\equiv {\langle{S_f}| A_s|{S_i}\9}/{\langle S_f\vert S_i\rangle}$.

    The key insight here is that since $\langle S_f\vert S_i\rangle$ can be a small number, the measurement of $q$ is influenced by
   a large multiplicative factor $\mathcal{A}_w$. A subsequent measurement of the meter reveals the desired parameter $q$. Trapped atoms are potentially promising system to implement this scheme~\cite{us-spin}.

   The simplest model of such a set-up consists of two
species of spins interacting with each other via a simple
exchange force, subject to the additional terms implied above, 
\begin{multline}
    H= J(\sigma_{1}^{S}\otimes \sigma_{1}^{M})+ \frac{\hbar g}{2c}\Big(h_{x}(\sigma_{1}^{S}+\sigma_{1}^{M})+
    \\(h_{z}+1)(\sigma_{3}^{S}+\sigma_{3}^{M}) + h_{y}(\sigma_{2}^{S}+\sigma_{2}^{M})\Big),\label{eq:1}
\end{multline}
which we write as $H = (\hbar\lambda/t)H_{0} + H_{1}$ where $H_{1}$ is the term proportional to $g$ and $\lambda = Jt/\hbar$ for convenience,
 with $h_{i}=-c\omega_{i}/g$. Analysis of the unitary evolution that is followed by post-selection indicates that for realistic parameter values the inertial and gravitational effects are within the
 sensitivity range of the optical magnetometery \cite{us-spin}.
 
 \begin{acknowledgements}
The work of DRT is supported by the  grant   FA2386-17-1-4015 of AOARD. ARHS was supported by the Natural Sciences and Engineering Research Council
of Canada and the Dartmouth College Society of Fellows. We  thank  Costantino Agnesi, L.C. Kwek and Alex Ling for useful discussions. We  acknowledge the International
Laser Ranging Service (ILRS) for SLR data and software.
\end{acknowledgements}

\end{document}